\documentclass[aps,preprint,showpacs,superscriptaddress,groupedaddress]{revtex4-1}  
\usepackage{graphicx}  
\usepackage{dcolumn}   
\usepackage{bm}        
\usepackage{amssymb}   
\usepackage[sort&compress]{natbib}

\begin{document}

\title{Quasiballistc Heat Conduction in Transient Grating Spectroscopy}

\author{Chengyun Hua}
\author{Austin J. Minnich}%
 \email{aminnich@caltech.edu}
\affiliation{%
 Division of Engineering and Applied Science\\
 California Institute of Technology, Pasadena, California 91125,USA
}%

\date{\today}

\begin{abstract}

Transient grating (TG) spectroscopy is an important experimental technique to measure mean free path (MFP) spectra using observations of quasiballistic heat conduction. To obtain MFP spectra, the measurements must be interpreted within the framework of the frequency-dependent  Boltzmann transport equation (BTE), but previous solutions have restricted validity due to simplifying assumptions. Here, we analyze heat conduction in TG using a new analytical solution of the frequency-dependent BTE that accurately describes thermal transport from the diffusive to ballistic regimes. We demonstrate that our result enables a more accurate measurement of MFP spectra and thus will lead to an improved understanding of heat conduction in solids. 

\end{abstract}

\maketitle 

\section{Introduction}

Quasiballistic heat conduction occurs if a temperature gradient exists over length scales comparable to phonon mean free paths (MFPs)\cite{Chen,Majumdar1993}. In this regime, local thermal equilibrium does not exist and Fourier's law is no longer valid. Presently, quasiballistic transport is under investigation due to its potential to infer information about the MFPs of thermal phonons\cite{Minnich2012}, knowledge of which is crucial to engineer thermal conductivity but remains unknown for most solids\cite{Dames2005,Minnich2009}. 

Quasiballistic transport was originally observed in macroscopic samples using heat pulse experiments\cite{Gutfeld1964} and later in silicon membranes using a microfabricated platform\cite{Sverdrup2001}. Nonlocal theories of heat conduction based on the Boltzmann transport equation (BTE) were introduced to describe the quasiballistic regime for phonons\cite{Mahan1988,Claro1989,Chen1996} and electrons\cite{Ezzahri2009}. Koh and Cahill reported modulation-frequency dependent thermal conductivities in a time-domain transient thermoreflectance (TDTR) experiment that they attributed to quasiballistic transport and suggested that the meansurements could be used to measure MFPs. Recently, quasiballistic transport has been observed in other experimental configurations\cite{Siemens2010,Minnich2011PRL,Malen2013NatComm,Johnson2013}. Minnich introduced a reconstruction technique that described how to quantitatively recover the MFP spectra from observations of quasiballistic heat transfer\cite{Minnich2012}. 

One notable experimental method for observing quasiballistic transport is the transient thermal grating (TTG) technique\cite{Johnson2013,Maznev2011,Rogers1994}, in which the interference of two laser pulses creates a sinusoidal initial temperature profile with wavelength $\lambda$. The observed thermal decay yields information about the thermal properties of the material. Recent work has demonstrated that these measurements can also reveal MFPs if the grating wavelength is comparable to MFPs, but interpreting measurements using the reconstruction method introduced by Minnich requires a solution of the BTE. A previous work reported a modified "two-channel" model\cite{Maznev2011}, in which low and high frequency phonons are described by the BTE and heat equation, respectively, but the extent of the validity of this model is unclear. An analysis within the framework of the BTE\cite{Collins2013APL} has been recently reported but the analysis of the frequency-dependent BTE was solely numerical, complicating its use for the reconstruction method. 

Here, we analyze thermal transport in TTG using a new analytical solution to the frequency-dependent BTE that is valid over the full range of the ballistic to diffusive regimes. Our analysis demonstrates the existence of weakly and strongly quasiballistic transport regimes that are distinguished by the thermal decay time relative to the phonon relaxation times. We provide theoretical justification for the use of a modified diffusion theory to interpret observations of quasiballistic transport. Finally, we use our solution to derive a corrected suppression function that enables phonon MFP spectra to be measured more accurately. Our results will lead to a better understanding of phonon heat conduction in solids like thermoelectrics.

\section{Modelling}

Thermal transport in a TTG experiment, assuming only in-plane heat conduction, is described by the 1D frequency-dependent BTE\cite{Majumdar1993},
\begin{eqnarray}\label{eq:BTE}
\frac{\partial g_{\omega}}{\partial t} &+& \mu v_{\omega} \frac{\partial g_{\omega}}{\partial x} = -\frac{g_{\omega}+f_0(T_0)-f_0(T)}{\tau_{\omega}} \\
f_0(T) &=& \frac{1}{4\pi}\hbar \omega D(\omega) f_{BE}(T) \approx f_0(T_0)+\frac{1}{4\pi}C_{\omega}\Delta T
\label{eq:BEDist_Linearized}
\end{eqnarray}
where $g_{\omega} = f_{\omega}(x,t,\mu)-f_0(T_0)$ is the deviational distribution function, $f_0 = f_0(x,t)$ is the equilibrium distribution function, $\mu = cos(\theta)$ is the directional cosine, $v_{\omega}$ is the phonon group velocity, and $\tau_{\omega}$ is the phonon relaxation time.  Assuming a small temperature rise, $\Delta T = T - T_0$, relative to a reference temperature, $T_0$, the equilibrium distribution is proportional to $\Delta T$, as shown in Eq.~(\ref{eq:BEDist_Linearized}).  Here, $\hbar$ is the reduced Planck constant, $\omega$ is the phonon frequency, $D(\omega)$ is the phonon density of states, $f_{BE}$ is the Bose-Einstein distribution, and $C_{\omega} = \hbar\omega D(\omega)\frac{\partial f_{BE}}{\partial T}$ is the mode specific heat. The volumetric heat capacity is then given by $C = \int_0^{\omega_m}C_{\omega}d\omega$ and the Fourier thermal conductivity $k = \int_0^{\omega_m}k_{\omega}d\omega$, where $k_{\omega} = \frac{1}{3} C_{\omega}v_{\omega} \Lambda_{\omega}$ and $\Lambda_{\omega} = \tau_{\omega}v_{\omega}$ is the phonon MFP. To close the problem, energy conservation is used to relate $g_{\omega}$ to $\Delta T$, given by 
\begin{equation}
\int\int_0^{\omega_m} \left[\frac{g_{\omega}(x,t)}{\tau_{\omega}}-\frac{1}{4\pi}\frac{C_{\omega}}{\tau_{\omega}}\Delta T(x,t) \right]d\omega d\Omega = 0
\label{eq:EnergyConservation}
\end{equation}
where $\Omega$ is the solid angle in spherical coordinates and $\omega_m$ is the cut-off frequency. Note that summation over phonon branches is implied without an explicit summation sign whenever an integration over phonon frequency or MFP is performed. 

Since the initial temperature profile in TTG is sinusoidal, we can assume that both $g_{\omega}$ and $\Delta T$ are of the form $e^{iqx}$, where $q = 2\pi/\lambda$ is the grating wavevector. Substituting $g_{\omega} = \widetilde{g}_{\omega}(t,\mu)e^{iqx}$ and $\Delta T = \Delta \widetilde{T}(t)e^{iqx}$ into Eq.~(\ref{eq:BTE}) leads to a first order ODE for $\widetilde{g}_{\omega}(t)$, and its solution is given by
\begin{equation}
 \widetilde{g}_{\omega} = \frac{1}{4\pi}\frac{C_{\omega}}{\tau_{\omega}}\int_0^t e^{\gamma (t'-t)}\Delta \widetilde{T}(t')dt' +  \widetilde{g}_{\omega}(0) e^{-\gamma t}
\label{eq:DistSolutionToBTE} 
\end{equation}
where $\gamma = (1+iq\mu \Lambda_{\omega})/\tau_{\omega}$ and $\widetilde{g}_{\omega}(0)$ is assumed to be independent of $\mu$. 

Applying the initial condition, $\widetilde{g}_{\omega}(0) = \Delta \widetilde{T}(0)  C_{\omega}/(4\pi)$ and  substituting Eq.~(\ref{eq:DistSolutionToBTE}) into Eq.~(\ref{eq:EnergyConservation}), we get an semi-analytical expression for $\Delta \widetilde{T}(t)$:
\begin{equation}
\Delta \widetilde{T}(t) = \frac{\Delta \widetilde{T}(0)}{2\int_0^{\omega_m} \frac{C_{\omega}}{\tau_{\omega}}d\omega} \int_0^{\omega_m}\int_{-1}^1 \left[\frac{C_{\omega}}{\tau_{\omega}^2}\int_0^t e^{\gamma (t'-t)}\frac{\Delta \widetilde{T}(t')}{\Delta \widetilde{T}(0)}dt' + \frac{C_{\omega}}{\tau_{\omega}}e^{-\gamma t}\right]d\mu d\omega.
\label{eq:TemperatureSolution}
\end{equation}

Eq.~(\ref{eq:TemperatureSolution}) is typically solved numerically by discretizing the integrals and solving a dense linear system, a computationally expensive task. Collins \emph{et. al.}\cite{Collins2013APL} obtained an analytical solution by applying a Fourier transform to the grey form of this equation. Here, we extend the Fourier transform method to the frequency-dependent BTE by observing that the time integral in Eq.~(\ref{eq:TemperatureSolution}) has the form of a convolution and thus can be simplified in the frequency domain. By applying a Fourier transform, we are able to decouple the nonlocal effects and obtain the following closed-form expression for the unknown distribution function $\widetilde{g}_{\omega}$ and transient temperature $\Delta \widetilde{T}$:
\begin{eqnarray}\label{eq:FreqDomainDist}
\mathcal{F}[\widetilde{g}_{\omega}](\xi) &=& \frac{1}{4\pi}\frac{C_{\omega}}{\tau_{\omega}}\frac{\mathcal{F}[\Delta \widetilde{T}](\xi)}{\gamma-i\xi}+\frac{C_{\omega}}{4\pi}\frac{\Delta \widetilde{T}(0)}{\gamma-i\xi}\\
\label{eq:FreqDomainTemp}
\mathcal{F}[\Delta \widetilde{T}](\xi) &=& \frac{ \Delta  \widetilde{T}(0)\int^{\omega_m}_0 C_{\omega}\mathcal{A}(\xi)d\omega}{\int^{\omega_m}_0\frac{C_{\omega}}{\tau_{\omega}}[1-\mathcal{A}(\xi)]d\omega}
\end{eqnarray}
\begin{equation}
\label{eq:FourierTransform} 
\mathcal{A}(\xi) = \mathcal{F}[\tau_{\omega}^{-1}e^{-t/\tau_{\omega}}\text{sinc}(qv_{\omega}t)u(t)]= \frac{i}{2q\Lambda_{\omega}}\log\left(\frac{\tau_{\omega}\xi+q\Lambda_{\omega}+i}{\tau_{\omega}\xi-q\Lambda_{\omega}+i}\right)
\end{equation}
where $\mathcal{F}$ denotes Fourier transform, $u(t)$ is the step function and $\xi$ is the Fourier variable. The time-domain solution is obtained by inverse fast Fourier transform. Therefore, we have derived an analytical solution to the frequency-dependent BTE that is valid from the ballistic to the diffusive regimes, enabling a more rigorous understanding of thermal transport in TTG. 

We can gain insight into which parameters determine the transport regime from our solution. From Eq.~(\ref{eq:FourierTransform}), we identify two nondimensional parameters. One is the familiar phonon Knudsen number Kn$_{\omega} = q\Lambda_{\omega}$, which compares the phonon MFP with a characteristic length, in this case $1/q$. To identify the second parameter, we notice that $\xi^{-1}$ describes a time scale that we assign to be the characteristic thermal decay time $\Gamma$. We can therefore define a new non-dimensional parameter that we denote the transient number, given by $\eta_{\omega} = \tau_{\omega}/\Gamma$, which compares the phonon relaxation times with the thermal decay time $\Gamma$. 

Note that the two parameters are not completely independent. For example, as the grating wavelength decreases, the thermal decay time also decreases. In the diffusion regime the relationship is trivial but in the quasiballistic and ballistic regimes the relationship becomes much more complex. While the Knudsen number can in principle completely distinguish the transport regime, we find that the transient number is an additional convenient parameter by which to specify the regime, particularly for quasiballistic transport where the specific Knudsen number at which a transition occurs is not obvious.

Therefore, together, these two numbers completely specify the transport regime. In the diffusive limit, length and time scales are much larger than the phonon MFPs and relaxation times, respectively, corresponding to Kn$_{\omega} \ll 1$ and $\eta_{\omega} \ll 1$. In the ballistic regime, lengths and times are much smaller than MFPs and relaxation times, or Kn$_{\omega} \gg 1$ and $\eta_{\omega} \gg 1$. The two regimes are well-understood limits of the BTE\cite{Chen}. Here, we focus on the intermediate range of the two limits, the quasiballistic regime.

\begin{figure*}
\centering
\includegraphics[scale = 0.5]{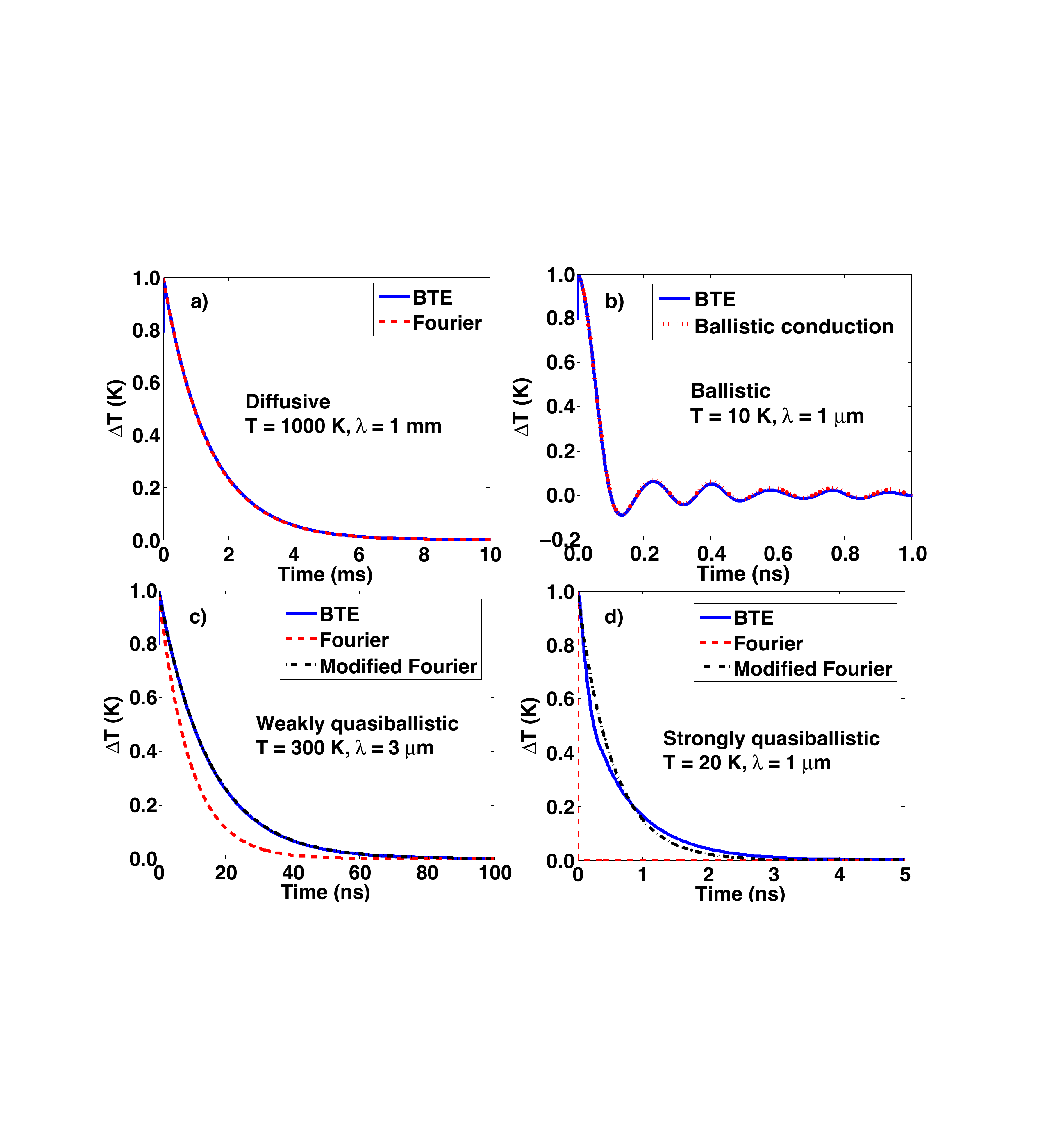}
\caption{Temperature decay curves $\Delta \widetilde{T}$ in the (a) diffusive limit, (b) ballistic limit, (c) weakly quasiballistic regime, and (d) strongly quasiballistic regime. The BTE solutions are given by the solid lines, the Fourier solution by the dashed lines, the ballistic conduction solution by the dotted line, and the Modified Fourier solution by the dash-dotted lines. }
\label{fig:TransientTemperature}
\end{figure*}

\section{Results}

To begin, we examine the transient temperature decay in the different regimes as shown in Fig.~\ref{fig:TransientTemperature}. We perform our calculations for crystalline silicon, using the experimental dispersion in the [100] direction and assuming the crystals are isotropic. The numerical details concerning the dispersion and relaxation times are given by Minnich's recent work\cite{Minnich2011PRB}.

\subsection{Diffusive and ballistic Limits}

We first confirm that our result correctly reproduces the diffusive and ballistic limits. Examining the limit of Eq.~(\ref{eq:FreqDomainTemp}) when both phonon relaxation times and MFPs are much smaller than their corresponding characteristic scales ($\eta_{\omega} \ll 1$ \& Kn$_{\omega}^2 \ll 1$), we find that the solution reduces to the Fourier solution and the thermal decay time $\Gamma = (q^2\alpha)^{-1}$, where $\alpha = k/C$ is the Fourier thermal diffusivity. Fig.~\ref{fig:TransientTemperature}a demonstrates that the BTE solution agrees with the Fourier solution in this limit. Similarly, at the ballistic limit ($\eta_{\omega} \gg 1$ \& Kn$_{\omega}^2 \gg 1$) shown in Fig.~\ref{fig:TransientTemperature}b, the transient temperature given by Eq.~(\ref{eq:FreqDomainTemp}) agrees with the ballistic solution of the BTE in which the relaxation times go to infinity. 

\subsection{Weakly Quasiballistic Regime}

We now examine the intermediate quasiballistic regime by allowing the MFPs to be comparable to or greater than the grating wavevector while requiring the thermal decay time to be much longer than relaxation times, Kn$_{\omega}^2 \sim 1$ but $\eta_{\omega} \ll 1$. We observe that the BTE solution does not agree with the Fourier's law solution, as shown in Fig.~\ref{fig:TransientTemperature}c. However, we observe that the shape of the temperature decay remains exponential, as in Fourier's law, but with a smaller thermal conductivity. We denote this regime the weakly quasiballistic regime, and the Fourier solution with a modified thermal conductivity as the modified Fourier solution. So far, the validity of the modified Fourier model to describe quasiballistic thermal transport is largely based on experimental observations\cite{Johnson2013}. The only theoretical approach to explain this observation was developed by Maznev \emph{et. al}. Their modified "two-channel" model assumes that the low-frequency phonons, which are analyzed by the BTE, only interact with the thermal reservoir of high-frequency phonons, which are analyzed by the diffusion equation. However, the extent of the validity of this assumption is not clear.

Here, we give a more rigorous explanation using our solution.  Under the assumption of Kn$_{\omega}^2 \sim 1$ and $\eta_{\omega} \ll 1$, the Taylor expansion of Eq.~\ref{eq:FourierTransform} around $\eta_{\omega} = 0$ gives
\begin{equation}\label{eq:AsymptoticFourierTransform}
\mathcal{A}(\xi) = i\eta_{\omega}\xi+\frac{\text{tan}^{-1}(\text{Kn}_{\omega})}{\text{Kn}_{\omega}} \sim 1.
\end{equation}
We observe that in the denominator of Eq.~(\ref{eq:FreqDomainTemp}), $1-\mathcal{A}(\xi) \sim \tau_{\omega}$ and the full asymptotic expression of $\mathcal{A}(\xi)$ should be used while in the numerator, $\mathcal{A}(\xi)$ can be approximated $1$. Therefore, Eq.~(\ref{eq:FreqDomainTemp}) asymptotically approaches the following form
\begin{equation}
\mathcal{F}[\Delta \widetilde{T}](\xi)\approx\frac{\Delta \widetilde{T}(0)}{q^2\alpha_{mod}-i\xi}
\label{eq:ReducedFourierTemp}
\end{equation}
\begin{equation}
k_{mod} = \int^{\omega_m}_0 k_{\omega}\left\{\frac{3}{(\text{Kn}_{\omega})^2}\left[1-\frac{\text{tan}^{-1}(\text{Kn}_{\omega})}{\text{Kn}_{\omega}}\right]\right\}d\omega
\label{eq:k_app}
\end{equation}
where $\alpha_{mod} = k_{mod}/C$ is the apparent thermal diffusivity and $k_{mod}$ is the modified thermal conductivity. Recognizing that Eq.~(\ref{eq:ReducedFourierTemp}) is simply the Fourier transform of an exponential decay, we find $\Delta \widetilde{T}(t) \approx \Delta \widetilde{T}(0)\text{exp}\left(-q^2\alpha_{mod} t\right)$. Thus, the formal solution of the BTE is equivalent to a modified diffusion theory with a modified thermal conductivity given by Eq.~(\ref{eq:ReducedFourierTemp}). The thermal decay time $\Gamma = (q^2\alpha_{mod})^{-1}$. We term this simplified solution the weak solution to the BTE, valid in the weakly quasiballistic regime. The modified thermal conductivity is the same expression given by Maznev \emph{et. al}\cite{Maznev2011}. 

Most recent experimental observations of quasiballistic transport have occurred in this weakly quasiballistic regime. For instance, in the TTG measurement of silicon membranes reported by Johnson \emph{et al.}\cite{Johnson2013}, the typical Kn$_{\omega} \approx 2.5$ and $\eta_{\omega} \sim O(0.01)$, based on the median thermal phonon MFP at the room temperature. Therefore, their measurements fall into the weakly quasiballistic regime and a modified Fourier solution should explain the results, in agreement with the experiment. 

\subsection{Strongly Quasiballistic Regime}

As the grating wavelength decreases, eventually the thermal decay becomes so fast that it is comparable to or greater than relaxation times such that $\eta_{\omega} \sim 1$ and Kn$_{\omega}^2 \sim 1$. Here, the assumption made in the modified "two-channel" model is not valid because some phonons in the thermal reservoir are now ballistic. For silicon, this regime occurs at small grating wavelength ($\lesssim 0.5\ \mu m$) or at cryogenic temperatures.  In this case, the BTE solution deviates from the exponential decay and can no longer be explained with any type of diffusion model as shown in Fig.~\ref{fig:TransientTemperature}d, and a full solution given by Eq.~(\ref{eq:FreqDomainTemp}) is necessary. We denote this regime the strongly quasiballistic regime. The equivalent decay time $\Gamma$ is given by $ \int_0^{\infty}\Delta T(t) dt/\Delta T(0)$, which reduces to the corresponding thermal decay times in the other two regimes above.

\subsection{Suppression Function}\label{sec:ThermalConductance}

We now seek to understand how the thermal length and time scales affect which phonons conduct heat in each regime. From our model, we can calculate the spectral thermal conductance, defined as the ratio of heat flux to the temperature difference
\begin{equation}
 \sigma_{\omega} = \frac{\int \int q_{\omega}(x,t)dxdt}{\int \int \Delta T(x,t)dxdt}
\label{eq:ThermalConductanceDefinition}
\end{equation} 
where $q_{\omega}(x,t) = \int g_{\omega}(x,t,\theta)v_{\omega}\cos(\theta) d\Omega$ is the spectral heat flux. In this way, we remove any spatial and temporal factors and can directly compare the heat flow induced by a unit temperature difference for each phonon mode. 

Substituting Eq.~(\ref{eq:DistSolutionToBTE}) into Eq.~(\ref{eq:ThermalConductanceDefinition}), we derive a general expression for the spectral thermal conductance
\begin{equation}
\sigma_{\omega} = \sigma_f \left\{\frac{3}{(\text{Kn}_{\omega})^2}\left[1-\frac{\text{tan}^{-1}(\text{Kn}_{\omega})}{\text{Kn}_{\omega}}\right]\left[\eta_{\omega}+1\right]\right\}
\label{eq:ThermalConductance}
\end{equation}
where $\sigma_f = k_{\omega}/(\lambda/2)$ is the Fourier thermal conductance. The term in the brackets, equal to the ratio of the BTE thermal conductance to the Fourier thermal conductance, was previously termed the suppression function $S(\text{Kn}_{\omega},\eta_{\omega})$ by Minnich\cite{Minnich2012}.

\begin{figure*}
\centering
\includegraphics[scale =1]{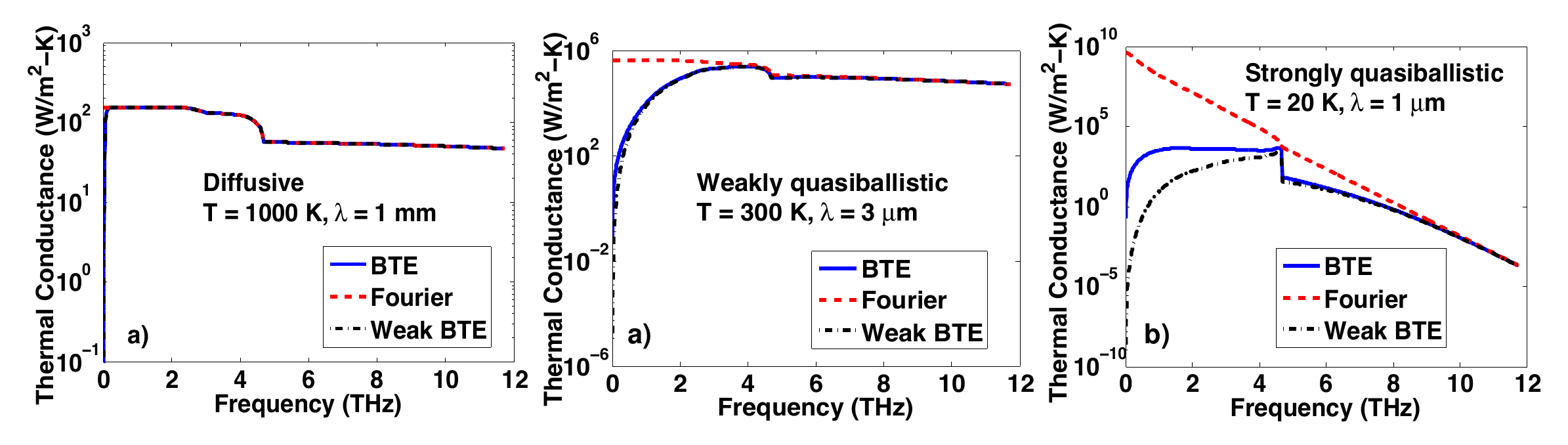}
\caption{Spectral thermal conductance $\sigma_{\omega}$ in the (a) weakly quasiballistic regime and (b) strongly quasiballistic regime. The BTE solutions are given by the solid lines, the Fourier solution by the dashed lines, and the weak BTE solution by the dash-dotted lines. }
\label{fig:ThermalConductance}
\end{figure*}

Now let us examine the thermal conductance in the two quasiballistic transport regimes discussed above, shown in Fig.~\ref{fig:ThermalConductance}. We compare the thermal conductance calculated by the Fourier's law, weak BTE and full BTE solutions. 

In weakly quasiballistic regime, where $\text{Kn}_{\omega}^2 \sim 1$ but $\eta_{\omega} \ll 1$ (Fig.~\ref{fig:ThermalConductance}a), $\eta_{\omega}+1 \rightarrow 1$ and Eq.~(\ref{eq:ThermalConductance}) reduces to $\sigma_{\omega} = \sigma_f S_{weak}(\text{Kn}_{\omega})$, where $S_{weak}(\text{Kn}_{\omega})$ is the same as the term in the brackets of Eq.~(\ref{eq:k_app}). In this regime, Fourier's law overpredicts the heat flux but the weak BTE solution still accurately describes the spectral heat distribution. From Fig.~\ref{fig:ThermalConductance}b, we see that the heat contribution from low frequency phonons is suppressed compared to the Fourier's law prediction. 

In the strongly quasiballistic regime (Fig.~\ref{fig:ThermalConductance}b), the weak BTE solution does not accurately explain the spectral conductance and we must instead use Eq.~(\ref{eq:ThermalConductance}). The full BTE solution predicts a more gradual suppression than the weak BTE solution for those low frequency phonons whose relaxation times are comparable to or greater than the thermal decay time $\Gamma$. This discrepancy is due to the correction term $\eta_{\omega}+1$, which approaches its maximum value at low frequencies and reduces the suppression effects. Rewriting $\eta_{\omega}$ into $\Lambda_{\omega}/(v_{\omega}\Gamma)$ in Eq.~(\ref{eq:ThermalConductance}), we find that our new suppression function decreases as $1/\Lambda_{\omega}$ in the long MFP limit, in agreement with the ballistic limit of the BTE\cite{Chen} while $S_{weak}(\text{Kn}_{\omega})$ predicts a steeper slope, $1/\Lambda_{\omega}^2$, which is inconsistent with the ballistic limit. Therefore, our new suppression function provides a more accurate prediction of the heat flux suppression over the entire spectrum of phonons compared to the approximate approaches in the literature\cite{Maznev2011,Collins2013APL}.

\begin{figure}
\centering
\includegraphics[scale = 0.32]{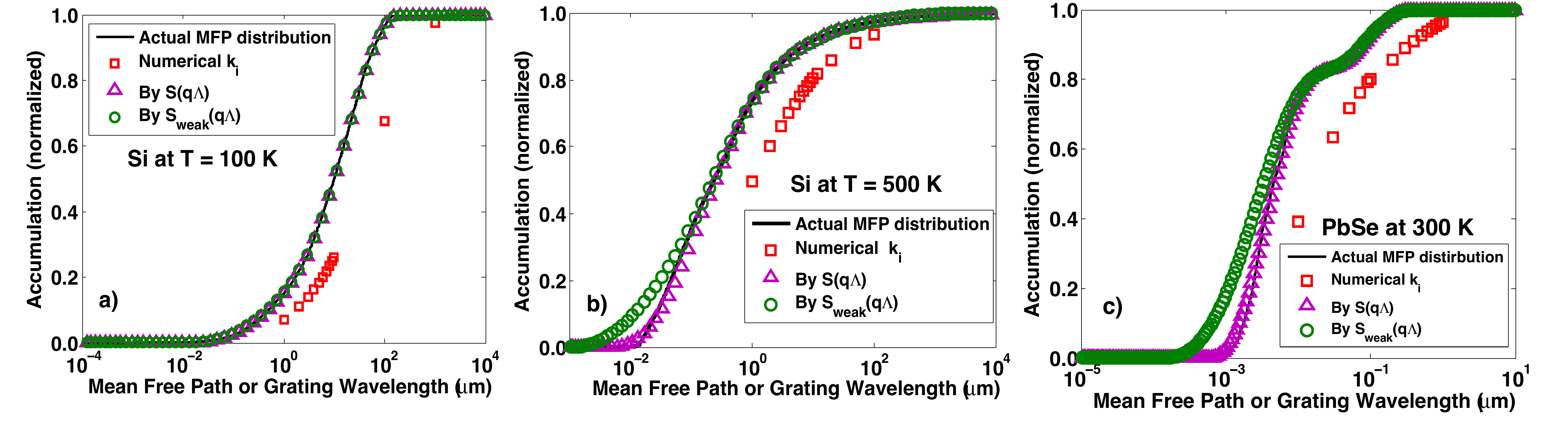}
\caption{Example MFP reconstructions for silicon at (a) 100 K and (b) 500 K and (c) PbSe  at 300 K using numerically simulated data. Plotted are the analytical MFP distribution (solid line), the numerical apparent thermal conductivities (squares), the reconstructed MFP distributions by the general suppression function (triangles) and by the weak suppression function (circles). The x-axis corresponds to the MFP for the distributions and to the grating wavelength for the thermal conductivity data.}
\label{fig:SuppressionFunction}
\end{figure}

\section{Application}

We now show the utility of these insights by demonstrating how our new suppression function may be used to more accurately measure MFP spectra. As proposed by Minnich\cite{Minnich2012}, the apparent thermal conductivities can be related to the MFP distribution by the equation $k_{app} = \int_0^{\infty}S(\Lambda_{\omega})f(\Lambda_{\omega})d\Lambda_{\omega}$ where $S(\Lambda_{\omega})$ is the suppression function, the phonon MFP $\Lambda_{\omega}$ is the independent variable and $f(\Lambda_{\omega})$ is the desired MFP distribution. If the apparent thermal conductivities are experimentally measured and the suppression function is known, then the MFP distribution can be reconstructed by solving the integral equation as an inverse problem. 

From our analysis, we have already derived the necessary suppression function $S(\text{Kn}_{\omega},\eta_{\omega})$ in Eq.~(\ref{eq:ThermalConductance}). However, this suppression function depends both on the independent variable, the phonon MFP $\Lambda_{\omega}$, as well as the unknown relaxation time $\tau_{\omega}$. To perform the reconstruction, $\Lambda_{\omega}$ should be the only unknown variable.

To overcome this problem, we rewrite $\tau_{\omega}$ into $\Lambda_{\omega}/v_{\omega}$ and assume that the phonon group velocity $v_{\omega}$ is equal to the average sound velocity $v_s$. This assumption is justified since for long MFP phonons for which the correction term is important, the group velocity of phonons is close to the sound speed, while for short MFP phonons this term is negligible and the choice of the velocity is irrelevant.  The reconstruction is also insensitive to the precise choice of the value of $v_s$. For example, for PbSe, changing $v_s$ from 2000 m/s to 1000 m/s causes only a 10 \% maximum error in the reconstructed MFP distribution. Using this approximation, the suppression function is only a function of the independent variable $\Lambda_{\omega}$. 

To demonstrate the inversion procedure using this new suppression function, we perform numerical experiments in which we obtain the modified thermal conductivities of Silicon and PbSe at different temperatures for different grating wavelengths from the temperature decay curves predicted by the BTE. These modified thermal conductivities, along with the suppression function, are then used as inputs for the reconstruction procedure. Fig.~\ref{fig:SuppressionFunction} shows the results for the MFP reconstruction using the general and weak suppression functions.

For materials with a MFP spectrum that is in the range of the experimental length scales, such as Silicon at 100 K, the measurements of the apparent thermal conductivities at different grating wavelengths span almost the entire range of phonon MFP spectrum. In this case, as shown in Fig.~\ref{fig:SuppressionFunction}(a), both the weak and new suppression functions yield satisfactory results. However, phonon MFPs vary by orders of magnitude and some part of the spectrum may be inaccessible to experiment. For example, the smallest MFPs of Silicon at 500 K are around 10 nm and the smallest MFPs of PbSe at room temperature are around 1 nm. These length scales are too small to be accessed with present experimental methods, meaning the MFP distribution at small length scales must be extrapolated from measurements at larger length scales. Such an extrapolation requires evaluating the suppression function at large values of the argument, precisely in the range where the correction term to the new suppression function is important.  As shown in Figs.~\ref{fig:SuppressionFunction}(b) and (c), our new suppression function yields more accurate results at short MFPs while the weak suppression function overpredicts the MFP distribution. 

\section{Summary}\label{sec:Summary}

We have analyzed thermal transport in TTG using a new analytical solution to the frequency-dependent BTE. We identify the thermal decay time relative to the relaxation times as a key nondimensional parameter that separate two quasiballistic transport regimes. If the thermal decay time is much larger than relaxation times, a modified diffusion theory is the formal solution of the BTE, providing theoretical justification for prior interpretations of experimental observations of quasiballistic transport. Further, we demonstrate how MFP spectra may be measured more accurately using our new suppression function. Our results will lead to a better understanding of phonon heat conduction in solids like thermoelectrics. 

\section*{Acknowledgement}
The authors would like to thank Kimberlee Collins and Gang Chen for useful discussions and for providing the PbSe data. This work was sponsored in part by Robert Bosch LLC through Bosch Energy Research Network Grant no. 13.01.CC11, by the National Science Foundation under Grant no. CBET 1254213, and by Boeing under the Boeing-Caltech Strategic Research \& Development Relationship Agreement.

\clearpage

\end{document}